\documentclass[aps,prl,reprint,groupedaddress,showpacs]{revtex4-1}

\usepackage{dcolumn}
\usepackage{amssymb}
\usepackage{graphicx}
\usepackage{bm}

\begin{document}


\title{Canonical Thermal Pure Quantum State}


\author{Sho Sugiura}
\email[]{sugiura@ASone.c.u-tokyo.ac.jp}
\author{Akira Shimizu}
\email{shmz@ASone.c.u-tokyo.ac.jp}
\affiliation{Department of Basic Science, 
University of Tokyo, 3-8-1 Komaba, Meguro, 
Tokyo 153-8902, Japan}


\date{\today}

\begin{abstract}
A thermal equilibrium state of a quantum many-body system
can be represented by 
a typical pure state, which we call a thermal pure quantum (TPQ) state. 
We construct the canonical TPQ state, 
which corresponds to the canonical ensemble of the conventional statistical mechanics. 
It is related to the microcanonical TPQ state,
which corresponds to the microcanonical ensemble,
by simple analytic transformations.
Both TPQ states give identical thermodynamic results, if both ensembles do, 
in the thermodynamic limit.
The TPQ states corresponding to other ensembles can also be constructed.
We have thus established the TPQ 
formulation of statistical mechanics, 
according to which 
all quantities of statistical-mechanical interest 
are obtained from a single realization of either TPQ state.
We also show that it 
has great advantages in practical applications. 
As an illustration, we study 
the spin-1/2 kagome Heisenberg antiferromagnet (KHA). 
\end{abstract}

\pacs{05.30.-d, 75.10.Jm, 05.70.Ce, 02.70.-c}

\maketitle

Statistical mechanics is conventionally described 
by the ensemble formulation, in which 
an equilibrium state is represented by 
a {\em mixed} quantum state.
Although its basic principles
are introduced in the microcanonical ensemble,
one can derive
other 
ensembles, such as the canonical ensemble. 
This  
makes statistical mechanics 
powerful and practical because
for most applications
the (grand) canonical ensemble is more convenient than 
the microcanonical one.

On the other hand, recent studies have shown that almost all 
{\em pure} quantum states
in a specified energy shell 
represent an equilibrium state 
\cite{Popescu,Lebowitz,SugitaJ,SugitaE,Reimann}. 
Generalizing this fact, 
we have defined a TPQ state
as a pure quantum state which represents an equilibrium state, 
and proposed 
a new formulation of statistical mechanics 
based on the TPQ state \cite{SS2012}.
Since the TPQ state proposed there 
corresponds to the microcanonical ensemble, 
we here call it the {\em microcanonical TPQ state}.
Then, natural questions arise:
Are there TPQ states which correspond to other 
ensembles?
How are they related to each other?

In this Letter, we construct a new type of TPQ state, 
the {\em canonical TPQ state}, 
which corresponds to the canonical ensemble. 
We 
show that it is related to the microcanonical TPQ state
by simple analytic transformations.
The TPQ states corresponding to other ensembles,
such as the grand canonical ensemble, 
can also be constructed in a similar manner.
These results establish
the new formulation of statistical mechanics, 
which enables one to obtain 
all quantities of statistical-mechanical interest 
from a {\em single} realization of a TPQ state.
This formulation is 
not only interesting as fundamental physics 
but also advantageous 
in practical applications
because one needs only to construct a single pure state
by just multiplying the Hamiltonian matrix to a random vector. 
The canonical TPQ state is particularly advantageous at low but finite temperature, whereas the microcanonical one is suitable for first-order phase transitions.
As an illustration, we study 
the KHA using the canonical TPQ state.

{\em Setup and definition -- } 
We consider a quantum system composed of $N$ sites or particles.
By using an effective model in the energy scale of interest \cite{discrete}, 
we describe the system with a Hilbert space $\mathcal{H}_N$ 
of dimension $\lambda^N$, 
where $\lambda=\Theta(1)$ \cite{Theta}.
[$\lambda=2$ for spin-$1/2$ systems.]
%
%
To take the thermodynamic limit, 
we use quantities per site, $\hat{h}{\equiv}\hat{H}/N$ and $u{\equiv}E/N$, 
where $\hat{H}$ denotes the Hamiltonian and $E$ the energy.
We do not write explicitly 
variables other than $u$ and $N$, such as a magnetic field.
{\em We assume 
that the ensemble formulation gives correct results,
which are consistent with thermodynamics
in the thermodynamic limit}.
[Here, the term `thermodynamics' is used 
in the sense of textbooks \cite{Callen,TD}.] 
This implies, e.g., that the entropy density $s(u)$ 
is a concave function and that the equivalence of ensembles holds.
A huge class of systems, 
{\em excluding} some models with long-range interactions \cite{TD,Thirring}, 
satisfy this condition.

Statistical mechanics 
treats 
macroscopic variables, which include 
`mechanical' and `genuine thermodynamic' ones.
{\em Mechanical variables}, such as energy and spin-spin
correlation functions, are the observables 
that are low-degree polynomials 
(i.e., their degree ${\leq}m$, where $m{=}o(N)$) 
of local operators. 
In order to exclude foolish operators (such as ${N^N}\hat{H}$), 
we assume that every mechanical variable $\hat{A}$ is
normalized such that 
$\|{\hat{A}}\|{\leq}KN^m$, 
where $K$ is a constant independent of $\hat{A}$ and $N$. 
By contrast, 
{\em genuine thermodynamic variables}, 
such as temperature and thermodynamic functions,
cannot be represented as such operators.
All genuine thermodynamic variables 
can be derived from the entropy function.

We have defined a TPQ state 
for the case where it 
has a random variable(s), 
as follows \cite{SS2012}.
A state $|\psi{\rangle}$ (${\in}\mathcal{H}_N$) is called a TPQ state 
if 
$
{\langle}\hat{A}{\rangle}^\psi_N
\stackrel{P}{\to}
{\langle}\hat{A}{\rangle}^{\rm ens}_N
$
uniformly for every mechanical variable $\hat{A}$ 
as $N \to \infty$.
Here, 
$
{\langle}\hat{A} \rangle^\psi_N
\equiv
{\langle}{\psi}|\hat{A}|{\psi}{\rangle}
/
\langle \psi | \psi \rangle
$,
$\langle{\cdot}\rangle^{\rm ens}_N$ is 
the ensemble average, and
`$\stackrel{P}{\to}$' denotes convergence in probability. 
That is, for an arbitrary positive number $\epsilon$ 
there exists a function $\eta_\epsilon(N)$ 
that vanishes in the thermodynamic limit 
and satisfies  
$
{\rm P} ( | 
{\langle}\hat{A}{\rangle}^\psi_N
-
{\langle}\hat{A}{\rangle}^{\rm ens}_N
|
{\geq}\epsilon
) 
\leq
\eta_\epsilon(N)
$
for every mechanical variable $\hat{A}$.
[${\rm P}(x)$ denotes the probability of event $x$.]
This means that for sufficiently large $N$ 
just a single realization of $|\psi{\rangle}$
gives the equilibrium values of all mechanical variables. 
%
%
Note that $|{\psi}\rangle$ is never obtained by `purification' of 
the density operator of the ensemble formulation, 
because $|{\psi}\rangle$ is not a vector in 
an enlarged Hilbert space but a vector in $\mathcal{H}_N$.

{\em Canonical TPQ state -- }
We have previously constructed the 
microcanonical TPQ state, which is specified by $(u,N)$ 
(i.e., independent variables are $u,N$) \cite{SS2012}. 
We now construct the canonical TPQ state, 
which is specified by $({\beta},N)$,
for finite inverse temperature ${\beta}=1/T$.

We take 
an {\em arbitrary} orthonormal basis of $\mathcal{H}_N$,
$\{|i\rangle\}_i$, which can be a trivial one 
such as a set of product states.
We also take 
random complex numbers $\{c_i\}_i$, 
which are drawn uniformly
from the unit sphere $\sum_i |c_i|^2=1$.
Then, 
$ 
|{\psi_0}\rangle 
\equiv
{\sum_i}{c_i}|i{\rangle}
$ 
is a random vector in $\mathcal{H}_N$.
Note that 
this construction of $|{\psi_0}\rangle$ is independent of the choice of 
$\{|i\rangle\}_i$. 
We will show that 
\begin{eqnarray}
|\beta,N \rangle 
\equiv 
\exp[-N \beta \hat{h} / 2] |\psi_0 \rangle
\label{Def: Phi}
\end{eqnarray} 
is the (unnormalized) canonical TPQ state specified by $({\beta},N)$.
%
We start with the rather obvious equality;
\begin{equation}
\overline{\langle \beta,N |\hat{A}| \beta,N \rangle}
\Big/
\overline{\langle \beta,N | \beta,N \rangle}
=
\langle \hat{A} \rangle^{\rm ens}_{\beta,N},
\label{Av<M>}
\end{equation}
%
where 
$\overline{\cdots}$ denotes the random average, 
${\langle}\hat{A}{\rangle}^{\rm ens}_{\beta,N} 
\equiv
{\rm Tr} \, [e^{-N \beta \hat{h}} \hat{A}]/Z(\beta,N)$, 
and $Z(\beta,N) \equiv {\rm Tr} \, e^{-N \beta \hat{h}}$. 
To investigate how fast 
$
{\langle}\hat{A}{\rangle}^{\rm TPQ}_{\beta,N}
\equiv
{\langle}{\beta,N}|\hat{A}|{\beta,N}\rangle
/
{\langle}{\beta,N}|{\beta,N}\rangle
$
converges to this value, 
we evaluate
$
D_N(A)^2
\equiv
\overline{
( 
{\langle}\hat{A}{\rangle}^{\rm TPQ}_{\beta,N}
		-{\langle}\hat{A}{\rangle}^{\rm ens}_{\beta,N}
)^2 
}
$ 
with the help of 
formulas in Ref.~\cite{Ullah}. 
Dropping smaller-order terms, we find \cite{SpMt} 
\begin{equation}
D_N(A)^2
\leq
{
\langle (\Delta \hat{A})^2 \rangle^{\rm ens}_{2\beta,N}
+
(\langle A \rangle^{\rm ens}_{2\beta,N} 
- \langle A \rangle^{\rm ens}_{\beta,N} )^2
\over \exp [2N\beta \{ f(1/2\beta ;N)-f(1/\beta ;N) \}]},
\label{Var<M>}
\end{equation}
%
where 
${\langle}(\Delta \hat{A})^2{\rangle}^{\rm ens}_{\beta,N}
{\equiv}{\langle}(\hat{A}-\langle{A}\rangle^{\rm ens}_{\beta,N})^2 
\rangle^{\rm ens}_{\beta,N}$, 
and
$f(T; N)$ is the free energy density. 
We here use $(T; N)$ instead of $(T,N)$ 
to indicate that $f(T; N)$ 
approaches the $N$-independent one, $f(T)$,
in the thermodynamic limit.
Since
$
\langle (\Delta \hat{A})^2 \rangle^{\rm ens}_{2\beta,N}
+
(\langle{A}\rangle^{\rm ens}_{2\beta,N} 
{-}{\langle}{A}{\rangle}^{\rm ens}_{\beta,N} )^2
\leq 5\| \hat{A} \|^2 \leq 5K^2 N^{2m}$,
and  
$f(1/2\beta;N)-f(1/\beta ;N) 
=
f(1/2\beta)-f(1/\beta) + o(1)
=  
\Theta(1) + o(1)$ 
at finite $\beta$ 
(because the entropy density 
$s = - \partial f/\partial T = \Theta(1)$),
we find 
$D_N(A)^2 \leq N^{2m}/e^{\Theta(N)}$, 
which becomes exponentially small with increasing $N$.

On the other hand, a generalized Markov's inequality yields, 
for arbitrary ${\epsilon}>0$,
\begin{equation}
{\rm P} \left( 
	\left|
\langle \hat{A} \rangle^{\rm TPQ}_{\beta,N}
	-\langle \hat{A} \rangle^{\rm ens}_{\beta,N}
	\right|
	\geq \epsilon
	\right)
\leq
D_N(A)^2 / \epsilon^2.
%
\label{P<}
\end{equation}
The right-hand side 
$\leq N^{2m} / \epsilon^2 e^{\Theta(N)}$, 
which 
vanishes exponentially fast with increasing $N$, 
for every mechanical variable $\hat{A}$.
Therefore, 
$
\langle \hat{A} \rangle^{\rm TPQ}_{\beta,N}
\stackrel{P}{\to}
\langle \hat{A} \rangle^{\rm ens}_{\beta,N}
$
uniformly,
which 
shows that $|\beta,N \rangle$ is the canonical TPQ state.
%

{\em Genuine thermodynamic variables --}
We now show that 
$|\beta,N \rangle$ also gives genuine thermodynamic variables correctly.
In the 
ensemble formulation, 
the partition function $Z(\beta,N)$ 
gives 
$f = -(1/\beta N) \ln Z$.
Similarly, in our formulation 
the squared norm 
of $|\beta,N \rangle$ 
gives $f$ 
as \cite{SpMt}
\begin{eqnarray}
&&
\overline{\langle \beta,N | \beta,N \rangle}
=
Z(\beta,N)/\lambda^N
=
\exp(-N \beta f)/\lambda^N,
\label{Av<>}
\\
&&
{\rm P} \left( 
	\left|
	\langle \beta,N | \beta,N \rangle
		/
		\overline{\langle \beta,N | \beta,N \rangle}
	-
	1
	\right|
	\geq \epsilon
	\right) 
\nonumber \\
&& \quad \qquad
\leq
1 
/ 
\epsilon^2 \exp [2N\beta \{ f(1/2\beta ;N)-f(1/\beta ;N) \}].
%
\quad
\label{Var<>}
\end{eqnarray}
Therefore, 
$
\langle \beta,N | \beta,N \rangle
\stackrel{P}{\to}
Z(\beta,N)/\lambda^N
$,
and
a single realization of 
the canonical TPQ state gives $f$, 
with exponentially small probability of error, 
by 
\begin{equation}
- \beta f(1/\beta ;N)
=
{1 \over N} 
\ln \langle \beta,N |\beta,N \rangle + \ln \lambda.
\label{ln<>=f}
\end{equation}
All genuine thermodynamic variables can be calculated from $f$ \cite{alsoMech}.
Furthermore, using $f$ obtained from this formula,
one can estimate the upper bounds of errors 
from formulas (\ref{Var<M>}), (\ref{P<}) and (\ref{Var<>}). 
In this sense, our formulas 
are almost self-validating ones. 
This property is particularly useful in practical applications
because one can confirm the validity of the results
without comparing them with results of other methods.

Note that the canonical ensemble and the canonical TPQ state
give the correct results {\em only when $N$ is large enough}.
For a system with small $N$, which interacts with a heat bath, 
{\em neither} gives the correct results in general, because 
interaction with the bath is non-negligible for small $N$.
To obtain the correct results for small $N$, 
one must apply {\em either} theory to a large system that includes 
the small system. 
Nevertheless, one might want results of the canonical ensemble
even for small $N$. 
In such a case, 
one can use Eqs.~(\ref{Av<M>}) and (\ref{Av<>})
by averaging over many realizations, 
because they exactly hold for {\em all} $N$.
%

We can show similar results for the unnormalized 
microcanonical TPQ state,
\begin{equation}
|k \rangle 
\equiv 
(l-\hat{h})^k |\psi_0 \rangle
\quad (k = 0, 1, 2, \cdots),
\label{mcTPQ}
\end{equation}
which represents the equilibrium state 
at the energy density 
$u_{k} = \langle k| \hat{h}| k \rangle/
\langle k| k \rangle$.
Here, 
$l$ is an arbitrary constant such that 
$l \geq \max \{$eigenvalue of $\hat{h} \}$ \cite{SS2012}. 
The squared norm $Q_k \equiv \langle k| k \rangle$
gives 
the entropy density $s(u; N)$ by \cite{better_s}
\begin{equation}
s(u_k; N)
=
{1 \over N} \ln Q_k - {2k \over N} \ln (l-u_k) + \ln \lambda + O(1/N).
\label{s=lnQ}
\end{equation}
This formula gives $s$ more directly
than the formula of Ref.~\cite{SS2012},
in which $s$ was obtained by integrating $\beta$.


{\em Equivalence and analytic relations-- }
In the ensemble formulation, the equivalence of ensembles
is important.
Also in our formulation, 
the canonical and the microcanonical TPQ states are
equivalent.
In fact, Eqs.~(\ref{Var<>}) and (\ref{s=lnQ})
show that both TPQ states give the correct thermodynamic
functions in the thermodynamic limit.
%
Since $\beta f(1/\beta)$ ($= \beta f(1/\beta;\infty)$) and
$s(u)$ ($=s(u;\infty)$) are equivalent (because they are 
related by the Legendre transformation), 
so are both TPQ states.
%
%

We can go one step further: 
These TPQ states 
are by themselves related by simple 
analytic transformations \cite{SpMt}.
To see this, we expand the exponential function of 
$e^{N \beta l/ 2} |\beta,N \rangle 
=
e^{N \beta (l -\hat{h}) / 2} |\psi_0 \rangle
$
%
%
as 
\begin{equation}
e^{N \beta l/ 2} |\beta,N \rangle 
=
\sum_{k=0}^{\infty} {(N \beta/2)^k \over k!} |k \rangle
=
\sum_{k=0}^{\infty} R_k |\psi_k \rangle.
\label{Phi=sumpsi_k}
\end{equation}
%
%
%
Here, 
$|\psi_k \rangle \equiv (1/\sqrt{Q_k}) |k \rangle$
is the normalized microcanonical TPQ state \cite{SS2012}, 
and 
$ 
R_k \equiv (N \beta/2)^k \sqrt{Q_k} /k!.
$ 
%
We can prove that 
the coefficient $R_k$ 
takes the maximum 
at $k^*$ ($=\Theta(N)$) such that 
\begin{equation}
\beta = \beta(u^*_{k^*}; N) + O(1/N),
\label{b=b(uk*)}
\end{equation}
where $u^*_k$ was defined in Ref.~\cite{SS2012},
and $\beta(u; N)$ is the inverse temperature 
of the equilibrium state specified by $(u, N)$,
i.e., 
$\beta(u; N) \equiv \partial s(u; N)/ \partial u$.
We can show that 
$u^*_{k^*} = \langle \psi_{k^*} |\hat{h}|\psi_{k^*} \rangle + O(1/N)$.
We can also prove that 
$R_k$ vanishes exponentially fast for $|k-k^*| \geq \Theta(N)$.
That is, $R_k$ 
has a sharp peak at $k^*$ and 
relevant terms in the sum 
are localized in the range $|k-k^*|=o(N)$. 
This means that 
$|\beta,N \rangle$ is almost composed of
$|\psi_k \rangle$'s whose temperature is close to $1/\beta$.
Such a natural and nice property is obtained because 
we have expanded $|\beta,N \rangle$ in powers not of $\hat{h}$
but of $(l-\hat{h})$.
Moreover, we can prove that the sum is uniformly convergent 
on any finite interval of $\beta$. 
That is, 
if one fixes arbitrarily the upper bound 
$\beta_{\rm max}$ ($>0$) of $\beta$ depending on one's purposes, 
then the sum is uniformly convergent for all $\beta$ 
such that $0 \leq \beta \leq \beta_{\rm max}$.
Because of this good convergence, 
we can obtain inversely the microcanonical TPQ state 
from the canonical one, e.g., by
%
\begin{eqnarray}
|k \rangle
&=&
\left( {2 \over N} \right)^k
{\partial^k  \over \partial \beta^k} 
e^{N \beta l/ 2} |\beta,N \rangle 
\bigg|_{\beta=0}.
\label{CantoMC}
\end{eqnarray}

{\em Various representations of equilibrium state -- }
The canonical density operator
$e^{-N \beta \hat{h}}/Z$ is invariant under time evolution.
$|\beta,N \rangle$ is also invariant
in the sense that it traverses various realizations of the same 
canonical TPQ state,
as can be seen by taking 
the energy eigenstates 
as the arbitrary basis $\{|i\rangle\}_i$ 
used in the construction of $| \psi_0 \rangle$.
%
Since almost all realizations of the TPQ state 
give identical results for macroscopic variables, 
as proved above, 
the TPQ state is {\em macroscopically} stationary
in consistent with thermodynamics.
%
Moreover, 
according to experience, any quantum state representing an equilibrium state
should be stable against weak external perturbations.
We can show using the results of Refs.~\cite{SM2002,SS2005} that 
the TPQ states (with an appropriate symmetry-breaking field(s)) 
do have such stability.
These facts support that 
an equilibrium state can be represented by 
various microstates, including 
$e^{-N \beta \hat{h}}/Z$ and
$|\beta,N \rangle$.


{\em Practical formulas -- }
It is practical to calculate 
$|\beta,N \rangle$
from $| \psi_k \rangle$'s 
through Eq.~(\ref{Phi=sumpsi_k}),
because the microcanonical TPQ states
$| \psi_1 \rangle, | \psi_2 \rangle, \cdots, | \psi_k \rangle$ can be obtained iteratively by simply 
multiplying $(l - \hat{h})$ with $| \psi_0 \rangle$ 
 $k$ times \cite{SS2012}.
Since $R_k$ has a sharp peak at $k^*$ (given by 
Eq.~(\ref{b=b(uk*)})), 
one can terminate the sum 
at a finite number $k_{\rm term}$.
It is sufficient to take $k_{\rm term}$ such that 
$k_{\rm term} - k^*_{\rm max} = \Theta(N)$,
where $k^*_{\rm max}$ is $k^*$ corresponding to $\beta_{\rm max}$.
Since we can show that $k^* = \Theta(N)$ for any finite $\beta$,
$k^*_{\rm max} = \Theta(N)$.
Hence, $k_{\rm term} = \Theta(N)$.
In this way, 
one can obtain $|\beta,N \rangle$ 
by multiplying $(l - \hat{h})$ repeatedly $\Theta(N)$ times.

All macroscopic variables can be calculated from the obtained 
$|\beta,N \rangle$.
One can also calculate them without 
obtaining $|\beta,N \rangle$ explicitly.
To see this, we note that all macroscopic variables 
can be obtained from 
$\langle \beta,N | \hat{A} |\beta,N \rangle$
and 
$\langle \beta,N | \beta,N \rangle$, 
as shown by formulas (\ref{P<}) and (\ref{ln<>=f}).
Since the latter is included in the former as the case of 
$\hat{A} = \hat{1}$, we consider the former.
From Eq.~(\ref{Phi=sumpsi_k}), 
$
\langle \beta,N |\hat{A}|\beta,N \rangle
=
e^{-N\beta l}
\sum_{k, k'} 
[(N \beta/2)^{k+k'} / k! k'!]
\langle k| \hat{A}| k' \rangle
$.
For the special case where $[\hat{A}, \hat{h}]=0$, this reduces to 
\begin{equation}
\langle \beta,N |\hat{A}|\beta,N \rangle
=
\{ \hat{A} \}'_{\beta,N},
\label{<M> practical}
\end{equation}
where
$
\{ \hat{A} \}'_{\beta,N}
\equiv
\sum_k
[(N\beta)^{2k} / (2k)!] \langle k|\hat{A}|k \rangle
+
\sum_k
[(N\beta)^{2k+1} / (2k+1)!]
\langle k|\hat{A}| k+1 \rangle.
$
%
%
Even when $[\hat{A}, \hat{h}] \neq 0$,
we can prove that 
Eq.~(\ref{<M> practical})
holds extremely well.
Specifically,
for 
$
\{ \hat{A} \}^{\rm TPQ}_{\beta,N}
\equiv
\{ \hat{A} \}'_{\beta,N}/\{ \hat{1} \}'_{\beta,N}
$
and
$E_N(A)^2
\equiv
\overline{(
\{ \hat{A} \}^{\rm TPQ}_{\beta,N}
-
\langle \hat{A} \rangle^{\rm ens}_{\beta,N}
)^2}
$,
we have
\begin{eqnarray}
&&
{\rm P} \left( 
	\left|
\{ \hat{A} \}^{\rm TPQ}_{\beta,N}
	-
	\langle \hat{A} \rangle^{\rm ens}_{\beta,N}
	\right|
	\geq \epsilon
\right)
\leq E_N(A)^2/\epsilon^2,
\label{P.practical}
\\
&& 
E_N(A)^2
\leq 
{\langle (\Delta \hat{A})^2 \rangle^{\rm ens}_{\beta,N}
\over \epsilon^2 \exp [N\beta \{ f(0;N)-f(1/\beta ;N) \}]},
\label{Var<star M>}
\end{eqnarray}
%
%
%
which show that 
$
\{ \hat{A} \}^{\rm TPQ}_{\beta,N}
\stackrel{P}{\to}
\langle \hat{A} \rangle^{{\rm ens}}_{\beta,N} 
$
exponentially fast and uniformly.
Formula (\ref{<M> practical}) is useful because 
one needs only to calculate 
$\langle k|\hat{A}|k\rangle$ and $\langle k|\hat{A}|k+1 \rangle$
for all $k\leq k_{\rm term}$
to obtain the results for {\em all} $\beta \leq \beta_{\rm max}$.

{\em The TPQ formulation of statistical mechanics -- } 
It is straightforward to extend 
the above theory to 
the TPQ states corresponding to other ensembles,
such as the grand canonical ensemble.
We have thus established the new formulation of statistical mechanics,
in the same level as the ensemble formulation.
It is summarized, for the microcanonical and 
canonical TPQ states, as follows.
%
Depending on the choice of independent variables, 
$(E, N)$ or $(\beta, N)$, 
one can use either state, 
because they give identical thermodynamic results.
A single realization of a TPQ state suffices 
for evaluating all quantities of statistical-mechanical interest. 
Moreover, one can estimate the upper bounds of errors
(which vanish as $N \to \infty$) 
by formulas 
(\ref{Var<M>}), (\ref{P<}), (\ref{Var<>}), (\ref{P.practical}) and (\ref{Var<star M>}).
The microcanonical and 
canonical TPQ states are transformed to each other 
by simple analytic relations, Eqs.~(\ref{Phi=sumpsi_k}) 
and (\ref{CantoMC}). 
Hence, getting either one implies getting both.
Using this fact, we have developed a practical 
formula (\ref{<M> practical}).

\begin{figure}
\begin{center}
\includegraphics[width=0.9\linewidth]{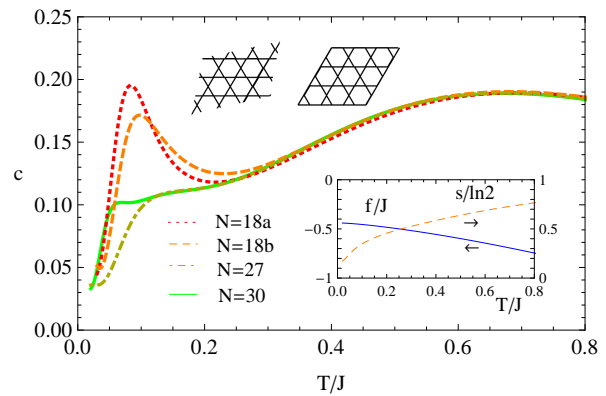}
\end{center}
\vspace{-6mm}
\caption{
$c$ vs. $T$ of the KHA. 
The shapes of clusters of $N=30$, $27$ and $18$a, $18$b
are shown in the left, right and in Ref.~\cite{N18}, respectively.
%
%
(Inset) $f$ and $s$ vs. $T$ for $N=30$. 
%
%
}
\label{fig:T}
\end{figure}

Since the TPQ formulation is much different from the ensemble formulation,
it will lead to deeper understanding of macroscopic quantum systems.
It is also 
unique and advantageous as a numerical method: 
(i) It computes expectation values for a {\em pure} quantum state, 
whereas the other methods compute 
those for a mixed state 
(such as $e^{-N \beta \hat{h}}/Z$). 
(ii) No limitation on models, 
i.e., applicable to any spatial dimensions and 
to complicated systems such as frustrated systems and fermion systems.
(iii) Applicable to {\em finite} temperature.
At finite $T$, 
there are an exponentially large number of states
in the corresponding energy shell. 
This reduces accuracy of many other methods.
By contrast, our method becomes {\em more accurate} as 
the number of relevant states increases, 
as explicitly shown 
by formulas
(\ref{Var<M>}), (\ref{Var<>}) and (\ref{Var<star M>}).
(iv) Almost {\em self-validating} because 
one can estimate the upper bounds of errors
from these formulas. 
(v) One can obtain TPQ states by simply multiplying 
$\hat{h}$ repeatedly $\Theta(N)$ times to a random vector.
This is much faster, e.g., than the numerical 
diagonalization (ND) of the full spectrum. 
[For example, it took only a few days to calculate all data 
in Fig.~\ref{fig:T} on a PC.] 
(vi) To obtain the results for all $\beta \leq \beta_{\rm max}$,
only two vectors, 
$| k \rangle$ and $| k+1 \rangle$, are required to store 
in the computer memory.
(vii) The orthogonality, $\langle k | k' \rangle =0$ for $k \neq k'$, is {\em not} necessary at all. 
This is advantageous to large-scale computations.

Regarding the choice between 
the canonical and microcanonical TPQ states,
one can use either 
depending on the purpose.
For example, if one is interested in 
a first-order phase transition at which the specific heat
$c = \partial u/ \partial T = (\partial T/ \partial u)^{-1}$ diverges
the microcanonical one is practically better,
because $T(u)$ is continuous (whereas $u(T)$ is discontinuous) 
through the transition \cite{TD,AS:phaserule}.
On the other hand, the canonical one is better 
when one studies low-temperature behavior of $c$,
because $\partial u(T)/ \partial T$ gets small 
($\partial T(u)/ \partial u$ diverges) as $c \to 0$.
%

{\em Application -- }
As an illustration, 
we study the KHA, 
which is a frustrated two-dimensional quantum spin system.
It was suggested that 
$c$ has double peaks 
at low temperature \cite{Elser}. 
However, the problem is still in dispute 
due to the complexity of the frustration and the finite size effect \cite{Elser, N18, Sindzingre, Isoda}.
We compute $c, f$ and $s$ for $N=18$-$30$, 
taking $k_{\rm term} = 2000$, 
for which 
the residual is evaluated to be less than $10^{-10} \%$ 
for 
$T\geq 0.02J$. 

In Fig.~\ref{fig:T} 
we plot $c$,  
obtained using
$
\partial \langle \hat{h} \rangle^{\rm ens}_{\beta,N} / \partial \beta 
= 
-\langle (\hat{h}-\langle \hat{h} \rangle^{\rm ens}_{\beta,N} )^2 \rangle^{\rm ens}_{\beta,N} 
\simeq
\{ \hat{\Delta}^2 \}^{\rm TPQ}_{\beta,N}$,
where 
$\hat{\Delta} \equiv \hat{h}-\{ \hat{h} \}^{\rm TPQ}_{\beta,N}$.
We have also calculated $c$ 
by using the difference method as
$\partial \langle \hat{h} \rangle^{\rm ens}_{\beta,N} / \partial \beta
\simeq 
(\{ \hat{h} \}^{\rm TPQ}_{\beta+\delta \beta,N} 
- \{ \hat{h} \}^{\rm TPQ}_{\beta,N})/\delta \beta$. 
The difference of these two numerical results is much smaller than the 
line width of the data in Fig.~\ref{fig:T}. 
For $N=27$ and $30$, for which ND has never been performed,
there is not a peak but a shoulder around $T=0.1J$, 
although
the finite size effect may still be  non-negligible. 

We also estimate the error caused from the random 
initial vector $|\psi_0 \rangle$ by using inequality (\ref{Var<star M>}). 
In the inset of Fig.~\ref{fig:T} we plot $f$ (left scale), 
which are calculated from Eq.~(\ref{ln<>=f}).
Using the results for $f$ and those for  
$\{ (\hat{\Delta}^2 - \{ \hat{\Delta}^2 \}^{\rm TPQ}_{\beta,N})^2 \}^{\rm TPQ}_{\beta,N}$, 
we find that 
the (normalized) standard deviation 
$D_N ( \hat{\Delta}^2 )
/\{ \hat{\Delta}^2 \}^{\rm TPQ}_{\beta,N}$
for $N=30$ is less than $1$\% down to $T=0.1J$. 
The error of $f$ itself is also estimated
to be less than $1$\% down to $T=0.1J$. 
Such a small error is attained 
because our method gets more accurate for larger entropy $Ns$ 
and 
the KHA has relatively large $s$ at low temperature 
due to the 
frustration effect \cite{Sindzingre}.
To see this quantitatively for $N=30$, 
we plot 
$s=(u - f)\beta$ 
in the inset of 
Fig.~\ref{fig:T} (right scale).
At $T=0.2J$ there remains $45$\% of the total entropy ($= N \ln 2$).
Such a large entropy makes $D_N ( \hat{\Delta}^2 )$ small.

Finally, to confirm the validity, 
we compute $c$ for $N=18$, 
for which the result of the ND is available \cite{N18}. 
%
%
For such a small cluster, 
the standard deviation estimated from inequality (\ref{Var<star M>}) 
is about $35$\% at $T=0.1J$. 
Hence, we have used Eq.~(\ref{Av<M>}) 
for $N=18$ only, 
taking average over $100$ realizations of the TPQ state.
The difference between our results ($18$a, $18$b) 
and those by the ND \cite{N18} is less than the 
line width of the data in Fig.~\ref{fig:T}.



\begin{acknowledgments}
We thank 
K. Asano, C. Hotta, K. Hukushima and H. Tasaki
for helpful discussions.
This work was supported by KAKENHI Nos.~22540407 and 23104707.
SS is supported by JSPS Research Fellowship No. 245328.
\end{acknowledgments}


\begin{thebibliography}{99}
\bibitem{discrete}
The Hilbert space 
of any quantum system is separable, in the irreducible representation. 
That is, its basis vectors are countable even when its dimension is infinite
as in particle systems.
Furthermore, one can take the basis vectors 
$|1 \rangle, |2 \rangle, \cdots$ 
in such a way that $\langle \nu | \hat{H} |\nu \rangle$ 
increases with increasing $\nu = 1, 2, \cdots$.
When the energy scale of interest is $E \ll E_{\rm max}$ ($\propto N$),
one can truncate the basis to 
$|1 \rangle, |2 \rangle, \cdots, |\nu_{\rm max} \rangle$,
where $\langle \nu_{\rm max} | \hat{H} |\nu_{\rm max} \rangle \sim E_{\rm max}$
(and $\ln[\nu_{\rm max}] \propto N$).
One thus obtains an effective Hilbert space of finite 
dimension $\nu_{\rm max}$.
Note however that 
Eqs.~(\ref{Def: Phi})-(\ref{ln<>=f}) hold even when $\lambda \to \infty$.
\bibitem{Theta}
For $f(N), g(N) > 0$, $f=\Theta(g)$ denotes that 
$^\exists a>0, ^\exists b >0, ^\exists N_0$ such that 
$a g(N) \leq f(N) \leq b g(N)$ for all $N >N_0$.
\bibitem{alsoMech}
Although mechanical variables can also be obtained from $f$,
formula (\ref{P<}) is often more convenient for them.
\end{thebibliography}

\end{document}